# Geospatial Big Data Handling with High Performance Computing: Current Approaches and Future Directions

***Zhenlong Li***

*Geoinformation and Big Data Research Laboratory, Department of Geography, University of South Carolina, Columbia, South Carolina*

zhenlong@sc.edu

**Abstract:** Geospatial big data plays a major role in the era of big data, as most data today are inherently spatial, collected with ubiquitous location-aware sensors. Efficiently collecting, managing, storing, and analyzing geospatial data streams enables development of new decision-support systems and provides unprecedented opportunities for business, science, and engineering. However, handling the "Vs" (volume, variety, velocity, veracity, and value) of big data is a challenging task. This is especially true for geospatial big data, since the massive datasets must be analyzed in the context of space and time. High performance computing (HPC) provides an essential solution to geospatial big data challenges. This chapter first summarizes four key aspects for handling geospatial big data with HPC and then briefly reviews existing HPC-related platforms and tools for geospatial big data processing. Lastly, future research directions in using HPC for geospatial big data handling are discussed.

Note: This article is accepted with minor revisions to be published in the book *High Performance Computing for Geospatial Applications* (Springer).

## 1. Introduction

Huge quantities of data are being generated across a broad range of domains, including banking, marketing, health, telecommunications, homeland security, computer networks, e-commerce, and scientific observations and simulations. These data are called *big data*. While there is no consensus on the definition of big data (Ward and Barker, 2013; De Mauro et al., 2015), one widely used definition is: “datasets whose size is beyond the ability of typical database software tools to capture, store, manage, and analyze” (Manyika et al., 2011, p.1).

*Geospatial big data* refers to a specific type of big data that contains location information. Location information plays a significant role in the big data era, as most data today are inherently spatial, collected with ubiquitous location-aware sensors such as satellites, GPS, and environmental observations. Geospatial big data offers great opportunities for advancing scientific discoveries across a broad range of fields, including climate science, disaster management, public health, precision agriculture, and smart cities. However, what matters is not the big data itself but the ability to efficiently and promptly extract meaningful information from it, an aspect reflected in the widely used big data definition provided above. Efficiently extracting such meaningful information and patterns is challenging due to big data’s 5-V characteristics—volume, velocity, variety, veracity, value (Zikopoulos and Eaton, 2011; Zikopoulos et al., 2012; Gudivada et al., 2015) —and geospatial data’s intrinsic feature of space and time. Volume refers to the large amounts of data being generated. Velocity indicates the high speed of data streams and that accumulation exceeds traditional settings. Variety refers to the high heterogeneity of data, such as different data sources, formats, and types. Veracity refers to the uncertainty and poor quality of data, including low accuracy, bias, and misinformation. For geospatial big data, these four Vs must be handled in the context of dynamic space and time to extract the ‘value’ from big data, which creates further challenges.

High performance computing (HPC) provides an essential solution to geospatial big data challenges by allowing fast processing of massive data collections in parallel. Handing geospatial big data with HPC can help us make quick and better decisions in time-sensitive situations, such as emergency response (Bhangale et al., 2016). It also helps us to solve larger problems, such as high-resolution global forest cover change mapping in reasonable timeframes (Hansen et al., 2013) and to achieve interactive analysis and visualization of big data (Yin et al., 2017).

This chapter explores how HPC is used to handle geospatial big data. Section 2 first summarizes four typical sources of geospatial big data. Section 3 describes the four key components, including data storage and management (section 3.1), spatial indexing (section 3.2), domain decomposition (3.3), and task scheduling (section 3.4). Section 4 briefly reviews existing HPC-enabled geospatial big data handling platforms and tools, which are summarized into four categories: general-purpose (section 4.1), geospatial-oriented (section 4.2), query processing (section 4.3), and workflow-based (section 4.4). Three future research directions for handling geospatial big data with HPC are suggested in section 5, including working towards a discrete global grid system (section 5.1), fog computing (section 5.2), and geospatial artificial intelligence (section 5.3). Lastly, section 6 summarizes the chapter.

## 2. Sources of Geospatial Big Data

Four typical sources of geospatial big data are summarized below.

- *Earth observations*

Earth observation systems generate massive volumes of disparate, dynamic, and geographically distributed geospatial data with in-situ and remote sensors. Remote sensing, with its increasingly higher spatial, temporal, and spectral resolutions, is one primary approach for collecting Earth observation data on a global scale. The Landsat archive, for example, exceeded one petabyte and contained over 5.5 million images several years ago (Wulder et al., 2016; Camara et al., 2016). As of 2014, NASA’s Earth Observing System Data and Information System (EOSDIS) was managing more than nine petabytes of data, and it is adding about 6.4 terabytes to its archives every day (Blumenfeld, 2019). In recent years, the wide use of drone-based remote sensing has opened another channel for big Earth observation data collection (Athanasis et al., 2018).

- *Geoscience model simulations*

The rapid advancement of computing power allows us to model and simulate Earth phenomena with increasingly higher spatiotemporal resolution and greater spatiotemporal coverage, producing huge amounts of simulated geospatial data. A typical example is the climate model simulations conducted by the Intergovernmental Panel on Climate Change (IPCC). The IPCC Fifth Assessment Report (AR5) alone produced ten petabytes of simulated climate data, and the next IPCC report is estimated to produce hundreds of petabytes (Yang et al., 2017; Schnase et al., 2017). Beside simulations, the process of calibrating the geoscience models also produces large amounts of geospatial data, since a model often must be run many times to sweep different parameters (Murphy et al., 2014). When calibrating ModelE (a climate model from NASA), for example, three terabytes of climate data were generated from 300 model-runs in just one experiment (Li et al., 2015).

- *Internet of Things*

The term *Internet of Things* (IoT) was first coined by Kevin Ashton in 1999 in the context of using radio frequency identification (RFID) for supply chain management (Ashton, 2009). Simply speaking, the IoT connects “things” to the internet and allows them to communicate and interact with one another, forming a vast network of connected things. The things include devices and objects such as sensors, cellphones, vehicles, appliances, and medical devices, to name a few. These things, coupled with now-ubiquitous location-based sensors, are generating massive amounts of geospatial data. In contrast to Earth observations and model simulations that produce structured multi-dimensional geospatial data, IoT continuously generates unstructured or semi-structured geospatial data streams across the globe, which are more dynamic, heterogeneous, and noisy.

- *Volunteered geographic information*

Volunteered geographic information (VGI) refers to the creation and dissemination of geographic information from the public, a process in which citizens are regarded as sensors moving “freely” over the surface of the Earth (Goodchild, 2017). Enabled by the internet, Web 2.0, GPS, and smartphone technologies, massive amounts of location-based data are being generated and disseminated by billions of citizen sensors inhabiting the world. Through geotagging (location sharing), for example, social media platforms such as Twitter, Facebook, Instagram, and Flickr provide environments for digital interactions among millions of people in the virtual space while leaving “digital footprints” in the physical space. For example, about 500 million tweets are sent per day according to Internet Live Stats (2019); assuming the estimated 1% geotagging rate (Marciniec, 2017), five million tweets are geotagged daily.

# 3. Key Components of Geospatial Big Data Handling with HPC

## 3.1 Data storage and management

Data storage and management is essential for any data manipulation system, and it is especially challenging when handling geospatial big data with HPC for two reasons. First, the massive volumes of data require large and reliable data storage. Traditional storage and protective fault-tolerance mechanisms, such as RAID (redundant array of independent disks), cannot efficiently handle data at the petabyte scale (Robinson, 2012). Second, the fast velocity of the data requires storage with flexibility to scale up or out to handle the ever-increasing storage demands (Katal et al., 2013).

There are three common types of data storage paradigms in HPC: shared-everything architecture (SEA), shared-disk architecture (SDA), and shared-nothing architecture (SNA) (Figure 1). With SEA, data storage and processing are often backed by a single high-end computer. The parallelization is typically achieved with multi-cores or graphics processing units (GPUs) accessing data from local disks. The storage of SEA is limited to a single computer and thus cannot efficiently handle big data.

SDA is a traditional HPC data storage architecture that stores data in a shared system that can be accessed by a cluster of computers in parallel over the network. Coupled with the message passing interface (MPI) (Gropp et al., 1996), the SDA-based HPC enables data to be transferred from storage to the compute nodes and processed in parallel. Most computing-intensive geospatial applications used it prior to the big data era. However, SDA does not work well with big data, as transferring large amounts of data over the network quickly creates a bottleneck in the system (Yin et al., 2013). In addition, the shared disk is prone to become the single point failure of the system.

Shared-nothing architecture (SNA) is not a new paradigm. Stonebraker pointed out in 1986 that shared-nothing was a preferred approach in developing multiprocessor systems at that time. With SNA, the data are distributedly stored on the cluster computers, each locally storing a subset of the data. SNA has become the de facto big data storage architecture nowadays because: (1) it is scalable, as new compute nodes can be easily added to an HPC cluster to increase its storage and computing capacity, (2) each data subset can be processed locally by the computer storing it, significantly reducing data transmission over the network, and (3) the single point failure is eliminated since the computers are independent and share no centralized storage.

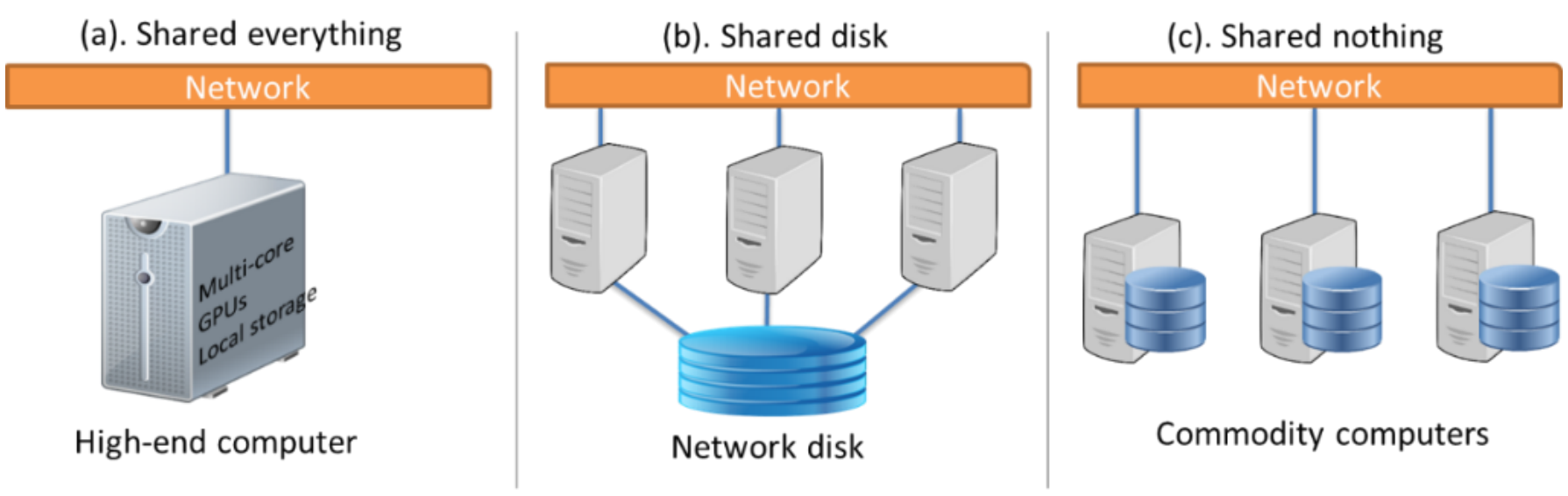

Figure 1. Illustration of different data storage architectures in HPC systems

One popular implementation of SNA is the Hadoop Distributed File System (HDFS) (Shvachko et al., 2010) —the core storage system for the Hadoop ecosystem. HDFS splits data into blocks and stores them across different compute nodes in a Hadoop cluster, so they can be processed in parallel. Like HDFS, most NoSQL (not only SQL) databases—including HBase (Vora, 2011), MongoDB (Abramova and Bernardino, 2013), and Google BigTable (Chang et al., 2008)—adopt SNA to store and manage big unstructured or semi-structured data. Since HDFS and NoSQL databases are not designed to store and manage geospatial data, many studies have been conducted to modify or extend these systems by integrating the spatial dimension (e.g., Wang et al., 2013; Zhang et al., 2014; Eldawy and Mokbel, 2015). Because the access patterns of a geospatial data partition (or block) are strongly linked to its neighboring partitions, co-locating the partitions that are spatially close with each other to a same computer node often improves data access efficiency in SNA (Fahmy, Elghandour, Nagi, 2016; Baumann et al., 2018).

### 3.2 Spatial indexing

With HPC, many processing units must concurrently retrieve different pieces of the data to perform various data processing and spatial analysis in parallel (e.g., clipping, road network analysis, remote sensing image classification). Spatial indexing is used to quickly locate and access the needed data, such as specific image tiles for raster data or specific geometries for vector data, from a massive dataset. Since the performance of the spatial index determines the efficiency of concurrent spatial data visits (Zhao et al., 2016), it directly impacts the performance of parallel data processing.

Most spatial indexes are based on tree data structures, such as the quadtree (Samet 1984), KD-tree (Ooi, 1987), R-tree (Guttman, 1984), and their variants. Quadtree recursively divides a two-dimensional space into four quadrants based on the maximum data capacity of each leaf cell (e.g., the maximum number of points allowed). A KD-tree is a binary tree often used for efficient nearest-neighbor search. An R-tree is similar to a KD-tree, but it handles not only point data but also rectangles such as geometry bounding boxes. As a result, R-trees and their variants have been widely used for spatial indexing (e.g., Xia et al., 2014; Wang et al., 2013). Especially focusing on geospatial big data, He et al. (2015) introduced a spatiotemporal indexing method based on decomposition tree raster data indexing for parallel access of big multidimensional movement data. SpatialHadoop uses an R-tree-based, two-level (global and local) spatial indexing mechanism to manage vector data (Eldawy and Mokbel, 2015) and a quadtree-based approach to index raster data (Eldawy et al., 2015).

The ability to store and process big data in its native formats is important because converting vast amounts of data to other formats requires effort and time. However, most indexing approaches for handling geospatial big data in an HPC environment (such as Hadoop) require data conversion or preprocessing. To tackle this challenge, Li et al. (2017) proposed a spatiotemporal indexing approach (SIA) to store and manage massive climate datasets in HDFS in their native formats (Figure 2). By linking the physical location information of node, file, and byte to the logical spatiotemporal information of variable, time, and space, a specific climate variable at a specific time, for example, can be quickly located and retrieved from terabytes of climate data at the byte level. The SIA approach has been extended to support other array-based datasets and distributed computing systems. For example, it was adopted by the National Aeronautics and Space Administration (NASA) as one of the key technologies in its Data Analytics and Storage System (DAAS) (Duffy et al., 2016). Based on SIA, Fu et al. (2018) developed an in-memory distributed computing framework for big climate data using Apache Spark (Zaharia et al., 2016). Following a concept similar to SIA, Li et al. (2018) developed a tile-based spatial index to handle large-scale LiDAR (light detection and ranging) point-cloud data in HDFS in their native LAS formats.

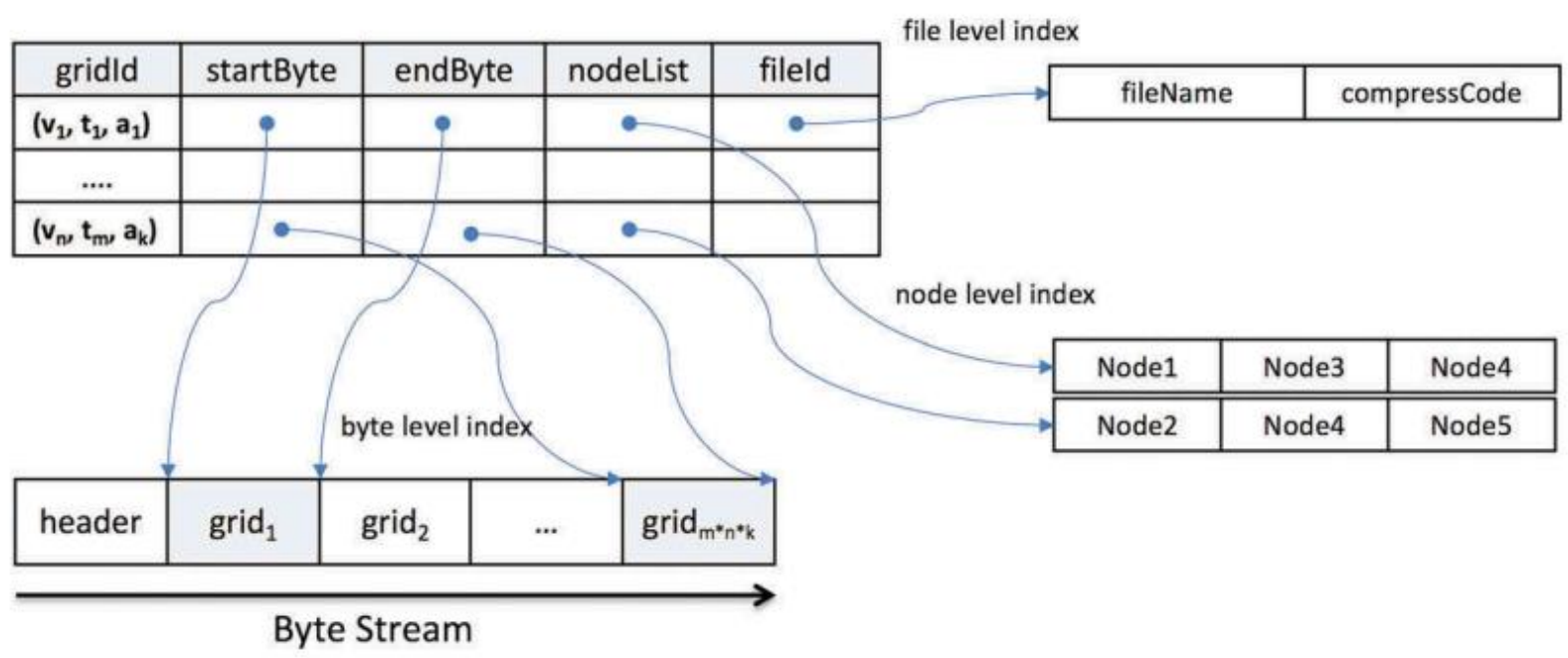

Figure 2. Illustration of the spatiotemporal indexing approach (Li et al., 2017)

### 3.3 Domain decomposition

Taking a divide-and-conquer approach, HPC first divides a big problem into concurrent small problems and then process them in parallel using multiple processing units (Ding and Densham, 1996). This procedure is called *decomposition*. Based on the problem to be solved, the decomposition will take one of three forms: domain decomposition, function decomposition, or both. Domain decomposition treats the data to be processed as the problem and decomposes them into many small datasets. Parallel operations are then performed on the decomposed data. Function decomposition, on the other hand, focuses on the computation, dividing the big computation problem (e.g., a climate simulation model) into small ones (e.g., ocean model, atmospheric model). We focus on domain decomposition here, as it is the typical approach used for processing geospatial big data with HPC.

Geospatial data, regardless of source or type, can be abstracted as a five-dimensional (5D) tuple < X, Y, Z, T, V>, where X, Y, Z denotes a location in three dimensional space, T denotes time, and V denotes a variable (spatial phenomenon), such as the land surface temperature observed at location X, Y, Z and time T. If a dimension has only one value, it is set to 1 in the tuple. For example, NASA's Modern-Era Retrospective analysis for Research and Applications (MERRA) hourly land surface data can be represented as <X, Y, 1, T, V> since there are no vertical layers. Based on this abstraction, domain decomposition can be applied to different dimensions of the data, resulting in different decompositions, such as 1D decomposition, 2D decomposition, and so on (Figure 3). The total number of subdomains produced by a domain decomposition equals the product of the number of slices of each domain.

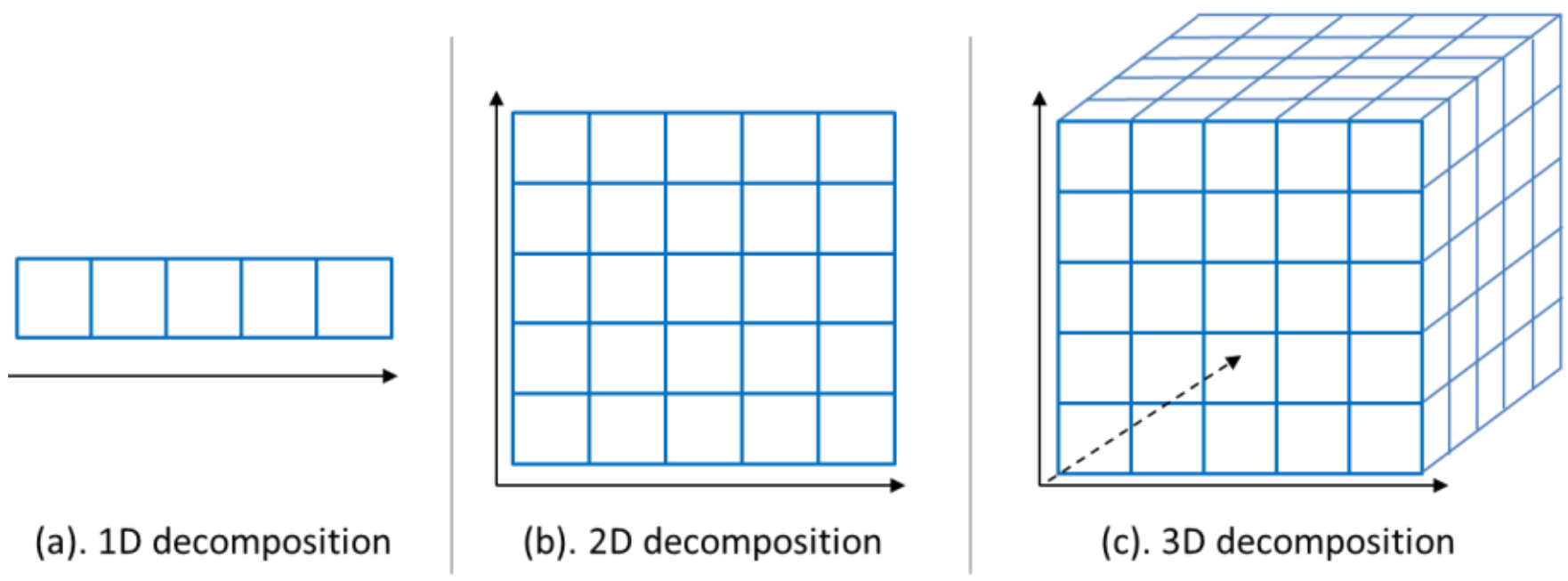

Figure 3. Illustration of domain decomposition. (a) 1D decomposition, decomposing any dimension of < X, Y, Z, T, V>; (b) 2D decomposition, decomposing any two dimensions of < X, Y, Z, T, V>; (c) 3D decomposition, decomposing any three dimensions of < X, Y, Z, T, V>.

Spatial decomposition occurs when data along the spatial dimensions <X, Y, Z> are decomposed. 2D spatial decomposition along <X, Y> often utilizes the regular grid or a quadtree-based approach, though irregular decomposition has also been used (Widlund, 2009; Guan, 2009). Wang and Armstrong (2003), for example, developed a parallel inverse-distance-weighted (IDW) spatial interpolation algorithm in an HPC environment using a quadtree-based domain decomposition approach. The quadtree was used to decompose the study area for adaptive load balancing. In a similar approach described by Guan, Zhang, and Clarke (2006), a spatially adaptive decomposition method was used to produce workload-oriented spatially adaptive decompositions. A more recent study by Li, Hodgson, and Li (2018) used a regular grid to divide the study area into many equal-sized subdomains for parallel LiDAR data processing. The size of the grid cell is calculated based on the study area size and available computing resources to maximize load balancing. Like 2D spatial decomposition, 3D spatial decomposition often uses a regular cube or octree-based approach to create 3D subdomains (Tschauner and Salinas, 2006). For example, Li et al. (2013) processed 3D environmental data (dust storm data) in parallel in an integrated GPU and CPU framework by equally dividing the data into 3D cubes.

Temporal decomposition decomposes data along the time dimension, which works well for time series data. Variable decomposition can be applied when a dataset contains many variables. For instance, MERRA land reanalysis data (MST1NXMLD) contains 50 climate variables that span from 1979 to the present with an hourly temporal resolution and a spatial resolution of 2/3 x 1/2 degree (Rienecker et al., 2011). In this case, the decomposition can be applied to the temporal dimension (T), the variable dimension (V), or both (T, V) (Li et al., 2015; Li et al., 2017).

When conducting domain decomposition, we need to consider whether dependence exists among the subdomains—in other words, whether a subdomain must communicate with others. For spatial decomposition, we need to check whether spatial dependence exists. For example, when parallelizing the IDW spatial interpolation algorithm using quadtree-based spatial decomposition, neighboring quads need to be considered (Wang and Armstrong, 2003). For some other operations, such as rasterizing LiDAR points, each subdomain can be processed independently without communicating with others (Li et al., 2018). For temporal decomposition, temporal dependence may need to be considered. For example, to extract the short- or long-term patterns from time series data requires considering temporal dependences in the decomposition (Asadi and Regan, 2019). Conversely, computing the global annual mean of an hourly climate variable does not require such consideration.

Knowing whether to consider dependence when decomposing data helps us design more efficient decomposition methods because avoiding unnecessary communications among subdomains often leads to better performance (Li et al., 2018). The problem of spatial dependence can be solved in multiple ways as summarized in Zheng et al., (2018). Spatial and temporal buffering can be used in domain decomposition to prevent communication with neighboring subdomains. For example, Hohl, Delmelle, and Tang (2015) used spatiotemporal buffers to include adjacent data points when parallelizing the kernel density analysis.

In addition to spatiotemporal dependence, the distribution of underlying data also needs special consideration for spatial and spatiotemporal domain decomposition because different data might pose different requirements for decomposition. For instance, while Hohl et al. (2018) decompose data that are distributed irregularly in all three dimensions, the data in Desjardins et al. (2018) are distributed irregularly in space, but regularly in time. As a result, different decomposition methods are used in the two examples for optimized performance.

### 3.4 Task scheduling

Task scheduling refers to distributing subtasks (subdomains) to concurrent computing units (e.g., CPU cores or computers) to be processed in parallel. Task scheduling is essential in HPC because the time spent to finish subtasks has a direct impact on parallelization performance. Determining an effective task schedule depends on the HPC programming paradigms and platforms (e.g., MPI-based or Hadoop-based), the problems to be parallelized (e.g., data-intensive or computation-intensive), and the underlying computing resources (e.g., on-premise HPC cluster or on-demand cloud-based HPC cluster). Regardless, two significant aspects must be considered to design efficient task scheduling approaches for geospatial big data processing: load balancing and data locality.

*Load balancing* aims to ensure each computing unit receives a similar (if not identical) number of subtasks for a data processing job, so that each finishes at the same time. This is important because in parallel computing, the job's finishing time is determined by the last finished task. Therefore, the number of subdomains and the workload of each should be considered along with the number of available concurrent computing units for load balancing. A load balancing algorithm can use static scheduling that either pre-allocates or adaptively allocates tasks to each computing unit (Guan, 2009; Shook et al., 2016). For example, Wang and Armstrong (2003) scheduled tasks based on the variability of the computing capacity at each computing site and the number of workloads used to partition the problem in a grid computing environment.

While most big data processing platforms (such as Hadoop) have built-in load balancing mechanisms, they are not efficient when processing geospatial big data. Hadoop-based geospatial big data platforms, such as GeoSpark (Yu, Wu, and Sarwat, 2015) and SpatialHadoop (Eldawy and Mokbel, 2015), often provide customized load balancing mechanisms that consider the nature of spatial data. For example, Li et al. (2017) used a grid assignment algorithm and a grid combination algorithm to ensure each compute node received a balanced workload when processing big climate data using Hadoop. When processing big LiDAR data, Li et al. (2018) calculated the number of subdomains to be decomposed based on the data volume and number of compute nodes in a cluster. In all cases, the subdomains should be comparably sized to better balance the load. In a cloud-based HPC environment, load balancing can also be achieved by automatically provisioning computing resources (e.g., add more compute nodes) based on the dynamic workload (Li et al., 2016).

*Data locality* refers to how close data are to their processing locations; a shorter distance indicates better data locality (Unat et al., 2017). Good data locality requires less data movement during parallel data processing and thus leads to better performance. Discussing data locality makes little sense in traditional HPC since it uses shared-disk architecture (section 2.1). A shared-disk architecture separates compute nodes and storage, thus requiring data movement. However, data locality is important for geospatial big data processing (Guo, Fox, and Zhou, 2012) because big data platforms (e.g., Hadoop) use shared-nothing storage; moving massive data among the compute nodes over the network is costly.

To archive data locality, the task scheduler is responsible for assigning a subdomain (data subset) to the compute node where the subdomain is located or stored. Thus, the task scheduler must know a subdomain's storage location, which can be realized by building an index to link data location in the cluster space to other spaces—geographic, variable, and file spaces. For instance, with a spatiotemporal index recording of the compute node on which a climate variable is stored, 99% of the data grids can be assigned to the compute nodes where the grids are stored, significantly improving performance (Li et al., 2017). In a LiDAR data processing study (Li et al., 2018), a spatial index was used to record a data tile's location in both the cluster and geographic spaces. Each subdomain was then assigned to the node where most of the tiles were stored. It is worth noting that besides load balancing and data locality, other factors such as computing and communication costs should also be considered for task scheduling.

## 4. Existing Platforms for Geospatial Big Data Handling with HPC

There are many existing platforms for handling geospatial big data with HPC. These offer various programming models and languages, software libraries, and application programming interfaces (APIs). Here I briefly review some of the popular platforms by summarizing them into four general categories.

### *4.1 General-purpose platforms*

General-purpose parallel programming platforms are designed to handle data from different domains. Open MPI, for example, is an open source MPI implementation for traditional HPC systems (Gabriel et al., 2004). Another open source HPC software framework is HTCondor (known as Condor before 2012), which supports both MPI and Parallel Virtual Machine (Thain, Tannenbaum, and Livny, 2005). Different from Open MPI and HTCondor, CUDA is a parallel computing platform designed to harness the power of the graphics processing unit (GPU) (Nvidia, 2011). GPU has a transformative impact on big data handling. A good example of how GPU enables big data analytics in the geospatial domain can be found in Tang, Feng and Jia (2015).

Entering the big data world, Hadoop, an open source platform, is designed to handle big data using a shared-nothing architecture consisting of commodity computers (Taylor, 2010). With Hadoop, big data is stored in the Hadoop distributed files system (HDFS) and is processed in parallel using the MapReduce programming model introduced by Google (Dean and Ghemawat, 2008). However, Hadoop is a batch processing framework with high latency and does not support real-time data processing. Apache Spark, an in-memory distributed computing platform using the same shared-nothing architecture as Hadoop, overcomes some of Hadoop's limitations (Zaharia et al., 2016).

### *4.2 Geospatial-oriented platforms*

As general-purpose platforms are not designed for handling geospatial data, efforts have been made to adapt existing parallel libraries or frameworks for them. Domain decomposition, spatial indexing, and task scheduling are often given special considerations when building geospatial-oriented programming libraries. One outstanding early work is GISolve Toolkit (Wang, 2008), which aims to enhance large geospatial problem-solving by integrating HPC, data management, and visualization in cyber-enabled geographic information systems (CyberGIS) environment (Wang, 2010; Wang et al., 2013). Later, Guan (2009) introduced an open source general-purpose parallel-raster-processing C++ library using MPI. More recently, Shook et al. (2016) developed a Python-based library for multi-core parallel processing of spatial data using a parallel cartographic modeling language (PCML).

In the big data landscape, an array of open source geospatial platforms has been developed based on Hadoop or Hadoop-like distributed computing platforms, including, for example, HadoopGIS (Wang et al., 2011), Geotrellis (Kini and Emanuele, 2014), SpatialHadoop (Eldawy and Mokbel, 2015), GeoSpark (Yu, Wu, and Sarwat 2015), GeoMesa (Hughes et al., 2015), EarthServer (Baumann et al., 2016), GeoWave (Whitby, Fecher and Bennight, 2017), and St_Hadoop (Alarabi, Mokbel, and Musleh, 2018). While not open source, Google Earth Engine (Gorelick et al., 2017) is a powerful and planetary-scale geospatial big data platform for parallel processing and analysis of petabytes of satellite imagery and other geospatial datasets.

### *4.3 Query processing*

Most general-purpose and geospatial-oriented programming libraries allow users to develop parallel data processing programs based on the APIs. Computer programming or scripting is generally needed, though some platforms offer high-level interfaces to ease development. Query processing falls into another category of big data processing that leverages structured query language for programming. Query processing, especially SQL-based, has gained noticeable popularity in the big data era, partly because it balances the usability and flexibility of a big data processing platform: more flexible than a static graphic user interface with fixed functions but less complicated than programming libraries (Li et al., 2019).

For raster data processing, the data can be naturally organized as data cubes (an array database), and traditional data cube operations—such as roll-up, drill-down, and slice—can be performed in parallel in an HPC environment. Examples of such platforms include RasDaMan (Baumann et al.,1999), SciDB (Cudré-Mauroux et al., 2009), and EarthDB (Planthaber, Stonebraker, and Frew, 2012). More recently, large scale raster data query processing has been investigated using Hadoop Hive and Apache Spark. Li et al. (2017), for example, introduced a query analytic framework to manage, aggregate, and retrieve array-based data in parallel with intuitive

SQL-style queries (HiveSQL). Based on the query analytical framework, an online scalable visual analytical system called SOVAS (Figure 4) was developed for query processing of big climate data using an extended-SQL as the query language (Li et al., 2019). Instead of using Hadoop, Hu et al. (2018) developed an in-memory big climate data computing framework based on the Spark platform that uses Spark SQL for query processing.

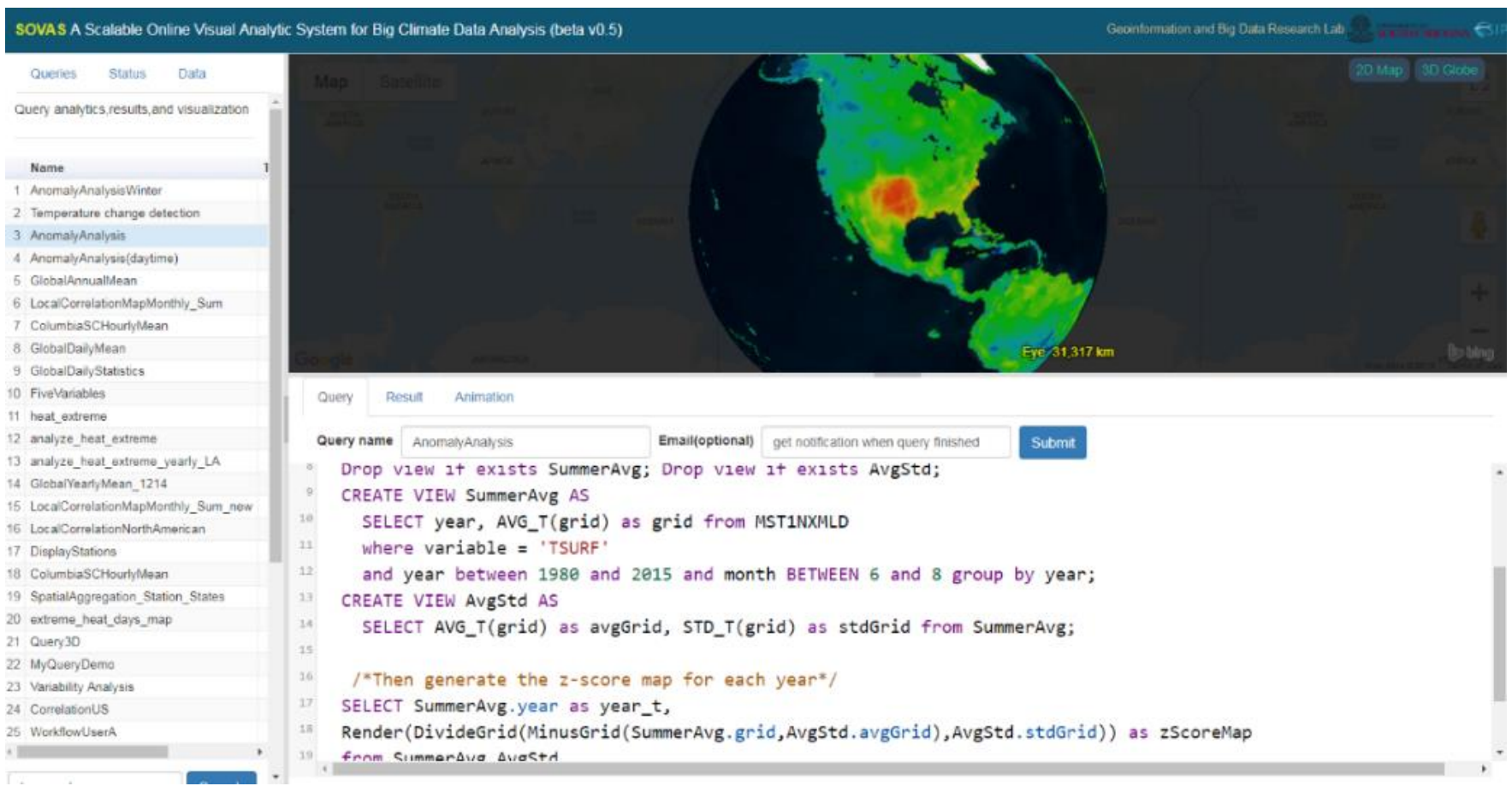

Figure 4. SQL-based query analytics of big climate data with SOVAS (https://gidbusc.github.io/SCOVAS)

PostGIS is a good example demonstrating how SQL works for vector data query processing (Ramsey, 2005). However, it falls short in handling geospatial big data due to its limited scalability. Esri tools for Hadoop (Esri, 2013) is one early effort to build a scalable big-vector data query processing framework based on Hadoop. In this framework, HiveSQL is the query language, and a suite of user-defined functions (UDFs) developed on top of the Esri Geometry API support various spatial operations, such as point-in-polygon and overlay. Later, Apache SparkSQL was adapted to develop a number of large-scale vector data query processing systems, such as GeoMesa SparkSQL (Kini and Emanuele, 2014), GeoSpark SQL (Huang et al., 2017), and Elcano (Engélinus and Badard, 2018). In contrast to these open source systems, Google BigQuery GIS offers a commercial tool that performs spatial operations using standard SQL to analyze big vector data (Google, 2019).

### 4.4 *Workflow-based systems*

Scientific workflow treats the data processing task as a pipeline consisting of a series of connected operations. For big data processing, an operation can be a parallel data processing task powered by HPC. There are many general-purpose scientific workflow systems developed to work in a distributed computing environment, including Kepler (Altintas et al., 2004), Triana (Taylor et al., 2005), Taverna (Hull et al., 2006), and VisTrails (Callahan et al., 2006). Since these workflow systems are not designed to work with geospatial data, efforts have been made to adapt them to build workflows for geospatial data processing (e.g., Jaeger et al., 2005; Zhang et al., 2006; Bouziane et al., 2008).

Geospatial service chaining is a service-based workflow approach for geospatial data processing in which each operation is provided as a web service (Yue, Gong, and Di, 2010; Gong et al., 2012). The web services used in the service chain are often based on the Open Geospatial Consortium's (OGC) standardized spatial web services for interoperability, including its Web Processing Service (WPS) for data processing, Web Feature Service (WFS) for vector data manipulation, Web Coverage Service (WCS) for raster data manipulation, and Web Mapping Service (WMS) for data visualization (Li et al., 2011). Over the past few years, studies have developed geospatial processing services running in the cloud-based HPC environment (Yoon et al., 2015; Tan et al., 2015; Baumann et al., 2016; Zhang et al., 2017; Lee and Kim, 2018).

A cloud-based HPC brings several advantages for geoprocessing workflow with big data, such as on-demand computing resource provision and high scalability. For example, Li et al. (2015) developed a cloud-based workflow framework for parallel processing of geospatial big data (Figure 5). In this framework, computing resources, such as Hadoop computing clusters and MaaS clusters (Li al., 2017), can be provisioned as needed when running the workflow.

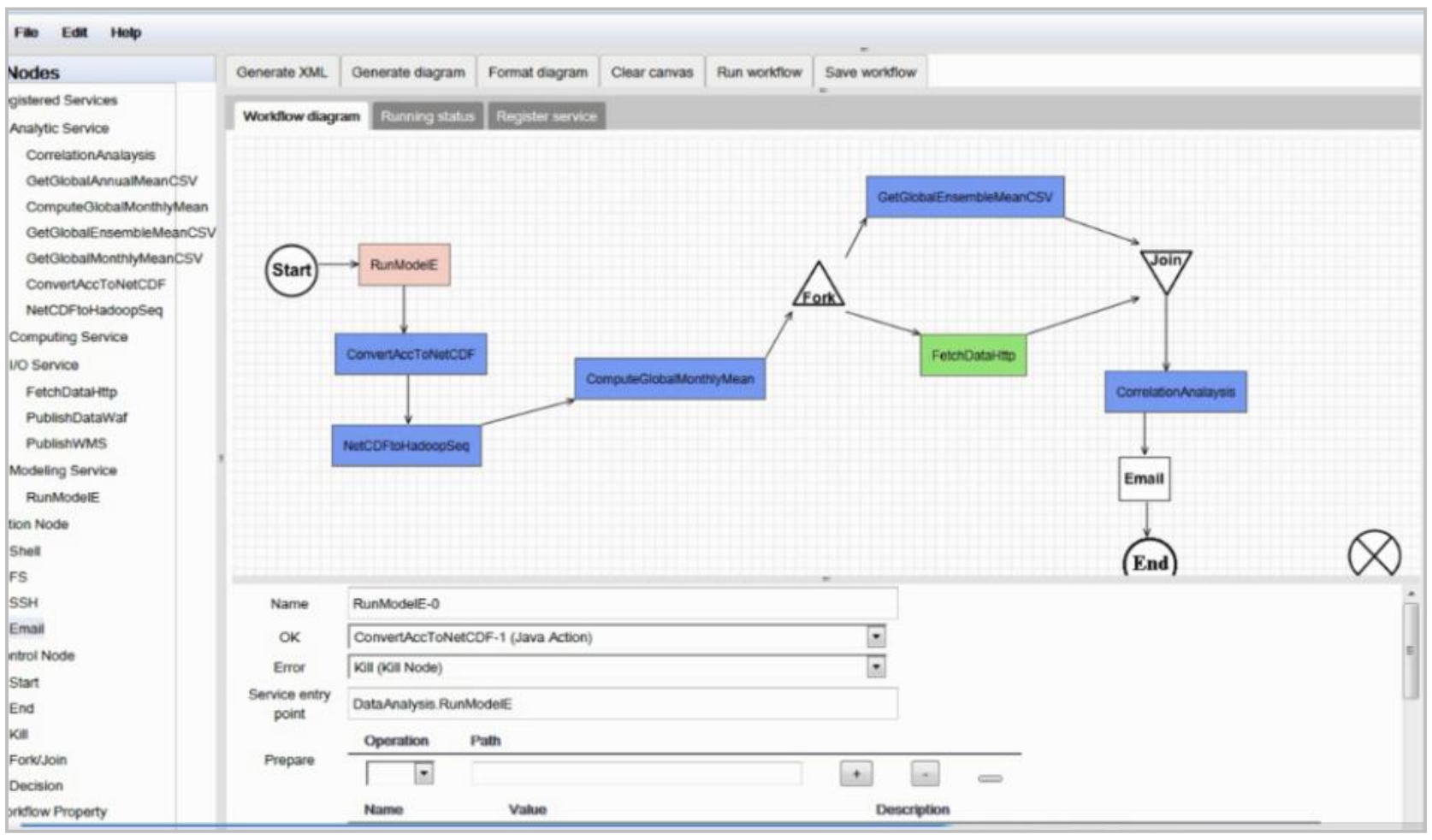

Figure 5. Geospatial big data handling using a cloud-based and MapReduce-enabled workflow

## 5. Directions for Further Research

### 5.1 Towards a discrete global reference framework with HPC

Heterogeneity has for a long time been a challenge in traditional geospatial data handling. Heterogeneity manifests in multiple aspects, including data collection approaches (e.g., remote sensing, land surveying, GPS), data models and formats (e.g., raster, vector), spatiotemporal scales/resolutions (e.g., from local to regional to world, from centimeters to meters to kilometers). Geospatial big data further creates heterogeneity through the ubiquitous location-based sensors collecting data from a broad range of sectors. Such heterogeneity makes it challenging to integrate and fuse geospatial big data with HPC. Most current HPC systems and studies handle a specific type of geospatial data with specific parallel algorithms, partly due to the lack of a referencing framework that can efficiently store, integrate, and manage the data in a way optimized for data integration and parallel processing.

While traditional coordinate systems (such as the system based on latitude and longitude) have been successful as a frame of reference, a relatively new framework called the discrete global grid system (DGGS) is believed to work better in managing and processing the heterogeneous geospatial big data associated with the curved surface of the Earth (Sabeur et al., 2019). DGGS represents "the Earth as hierarchical sequences of equal area tessellations on the surface of the Earth, each with global coverage and with progressively finer spatial resolution" (OGC, 2017). It aims to provide a unified, globally consistent reference framework to integrate heterogeneous spatial data—such as raster, vector, and point cloud—with different spatiotemporal scales and resolutions. The design of DGGS makes it natively suitable for parallel processing with HPC, as the data that it stores and manages has already been decomposed into discrete subdomains. However, currently most HPC-based spatial data processing research and tools remain based on traditional reference frameworks. Future research is needed to investigate spatiotemporal indexes, parallel algorithms, and big data computing platforms in the context of DGGS and HPC.

### 5.2 Towards fog computing with HPC

Fog computing is an emerging computing paradigm that resides between smart end-devices and traditional cloud or data centers (Iorga et al., 2017). It aims to process big data generated from distributed IoT devices (also called edge devices) in real time to support applications such as smart cities, precision agriculture, and autonomous vehicles. In traditional IoT architecture, the limited computing power of edge devices means the data they generate are directly uploaded to the cloud with no or very limited processing. This creates noticeable latency because the data are often far away from the cloud (poor data locality). Fog computing provides a middle computing layer – a cluster of fog nodes– between the edge devices and cloud. Since the fog nodes have more computing power and are close to the edge devices with low network latency (good data locality), edge device data can be quickly transferred to them for real-time filtering and processing. The filtered data can then be transferred to the cloud as needed for data mining and analysis using Hadoop-like systems, artificial intelligence, or traditional HPC.

IoT generates geospatial big data, thanks to the ubiquitous location-based sensors on edge devices. In this sense, real-time geospatial data processing is critical in fog computing. HPC should be researched and utilized in fog computing to deliver real-time responses for decision making (e.g., by an autonomous vehicle) from the following aspects: (i). Geospatial data processing in the cloud: As cloud computing plays an important role in fog computing, research on how to efficiently transfer data from edge devices to the cloud and to process geospatial data in parallel in a cloud environment is greatly needed. (ii). Geospatial data processing on the fog node: Since fog computing aims to provide real-time data processing, research is needed to design parallel computing algorithms and platforms that better utilize the embedded, mobile, and low-end fog node computers. (iii). Geospatial data processing in the fog cluster: Fog nodes are connected with a high-speed, low-latency network, which can form a high performance computing cluster. Unlike traditional computing clusters, such nodes might be mobile within a complex networking environment. For example, if autonomous cars are deployed as fog nodes, we could use those parked in a garage as a computing cluster. The challenges include, for example, how to efficiently form a computing cluster considering the spatial locations of fog nodes, how to use domain decomposition to assign the distributed edge devices to fog nodes, and how to develop smart scheduling algorithms to assign data processing tasks to appropriate nodes.

### 5.3 Towards geospatial artificial intelligence with HPC

Artificial intelligence (AI) is a computer science field that uses computers to mimic human intelligence for problem-solving (Minsky, 1961). Deep learning, a branch of machine learning in AI, has made significant progress in recent years with a broad range of applications, such as natural language processing and computer visions (Chen and Lin, 2014; LeCun, Bengio, and Hinton, 2015). Unlike traditional machine learning, in which parameters of an algorithm (e.g., support vector machine) are configured by experts, deep learning determines these parameters by learning the patterns in a large amount of data based on artificial neural networks.

Geospatial artificial intelligence (GeoAI) uses AI technologies like deep learning to extract meaningful information from geospatial big data (VoPham et al., 2018). GeoAI has had success across a broad range of applications, especially in remote sensing, such as image classification (Hu et al., 2015), object detection (Cheng et al., 2016), and land cover mapping (Kussul et al., 2017; Ling and Foody, 2019). While GeoAI is a promising solution for geospatial big data challenges, geospatial big data is likewise critical in training GeoAI's complex deep neural networks (DNNs) and is the catalyst that has stimulated deep learning advancements in recent years. As highlighted by Jeff Dean (2016) of the Google Brain team, an important property of neural networks is that results improve when using more data and computations to train bigger models. This is where high performance computing comes into play.

Tech giants such as Google, Microsoft, and IBM, have been leading the development of large-scale AI platforms that run on big computing clusters. Most current GeoAI research in the literature, however, is conducted on single-node computers or workstations using relatively small amounts of data to train the model. For example, Zhang et al. (2018) conducted an object-based convolutional neural network for urban land use classification based on only two 0.5 m resolution images of about 6,000 × 5,000 pixels. A recent review reveals that 95.6% of published research on remote sensing land-cover image classification covers less than 300 ha and uses small training sets (Ma et al., 2017). One potential reason is the lack of geospatial-oriented deep learning platforms available for academic research that support parallelization in a distributed environment. For example, DeepNetsForEO, an open source deep learning framework based on the SegNet architecture for semantic labeling of Earth observation images (Badrinarayanan et al., 2017; Audebert, Saux, and Lefèvre, 2018), only supports reading the entire training set into the computer memory, which is not scalable to large datasets.

More research, from the geospatial big data and engineering perspectives, is urgently needed to develop high-performance, scalable GeoAI frameworks and platforms that take full advantage of geospatial big data to build bigger and better models. This can be achieved by integrating general-purpose deep learning platforms, such as TensorFlow (Abadi et al., 2016), Caffe (Jia et al., 2014), and Apache SINGA (Ooi et al, 2015), with HPC technologies in the geospatial context, similar to adopting general-purpose big data platforms in Hadoop to handle geospatial big data. Specific research directions might include the development of efficient spatiotemporal indexing, domain decomposition, and scheduling approaches to parallelize a deep convolutional neural network in a distributed HPC environment.

## 6. **Summary**

Geospatial big data is playing an increasingly important role in the big data era. Effectively and efficiently handling geospatial big data is critical to extracting meaningful information for knowledge discovery and decision making, and HPC is a viable solution. This chapter began with a brief introduction of geospatial big data and its sources and then discussed several key components of using HPC to handle geospatial big data. A review of current tools was then provided from four different aspects. Lastly, three research directions were discussed in the context of HPC and geospatial big data.

HPC has been used for geospatial data handling for almost two decades (Armstrong, 2000; Clarke, 2003; Wang and Armstrong, 2003) and is becoming more important in tackling geospatial big data challenges. Geospatial big data, in turn, brings new challenges and opportunities to HPC. It is evident that the interweaving of geospatial big data, cloud computing, fog computing, and artificial intelligence is driving and reshaping geospatial data science. High performance computing, with its fundamental divide-and-conquer approach to solving big problems faster, will continue to play a crucial role in this new era.

## **References**

Abadi, M., Barham, P., Chen, J., Chen, Z., Davis, A., Dean, J., ... & Kudlur, M. (2016). Tensorflow: A system for large-scale machine learning. In 12th {USENIX} Symposium on Operating Systems Design and Implementation ({OSDI} 16) (pp. 265-283).

Abramova, V., & Bernardino, J. (2013, July). NoSQL databases: MongoDB vs cassandra. In *Proceedings of the international C**

*conference on computer science and software engineering* (pp. 14-22). ACM

Aji, A., Wang, F., Vo, H., Lee, R., Liu, Q., Zhang, X., & Saltz, J. (2013). Hadoop gis: a high performance spatial data warehousing system over mapreduce. Proceedings of the VLDB Endowment, 6(11), 1009-1020.

Alarabi, L., Mokbel, M. F., & Musleh, M. (2018). St-hadoop: A mapreduce framework for spatio-temporal data. GeoInformatica, 22(4), 785-813.

Altintas, I., Berkley, C., Jaeger, E., Jones, M., Ludascher, B., & Mock, S. (2004, June). Kepler: an extensible system for design and execution of scientific workflows. In Proceedings. 16th International Conference on Scientific and Statistical Database Management, 2004. (pp. 423-424). IEEE.

Armstrong, M P. 2000. Geography and computational science. Annals of the Association of American Geographers, 90(1): 146–56.

Asadi, R., & Regan, A. (2019). A Spatial-Temporal Decomposition Based Deep Neural Network for Time Series Forecasting. arXiv preprint arXiv:1902.00636.

Ashton, K. (2009). That 'internet of things' thing. RFID journal, 22(7), 97-114.

Athanasis, N., Themistocleous, M., Kalabokidis, K., & Chatzitheodorou, C. (2018, October). Big Data Analysis in UAV Surveillance for Wildfire Prevention and Management. In European, Mediterranean, and Middle Eastern Conference on Information Systems (pp. 47-58). Springer, Cham.

Audebert, N., Le Saux, B., & Lefèvre, S. (2018). Beyond RGB: Very high resolution urban remote sensing with multimodal deep networks. ISPRS Journal of Photogrammetry and Remote Sensing, 140, 20-32.

Badrinarayanan, V., Kendall, A., & Cipolla, R. (2017). Segnet: A deep convolutional encoder-decoder architecture for image segmentation. IEEE transactions on pattern analysis and machine intelligence, 39(12), 2481-2495.

Baumann, P., Dehmel, A., Furtado, P., Ritsch, R., & Widmann, N. (1998). The multidimensional database system RasDaMan. Acm Sigmod Record, 27(2), 575-577.

Baumann, P., Dehmel, A., Furtado, P., Ritsch, R., & Widmann, N. (1999, September). Spatio-temporal retrieval with RasDaMan. In VLDB (pp. 746-749).

Baumann, P., Mazzetti, P., Ungar, J., Barbera, R., Barboni, D., Beccati, A., ... & Campalani, P. (2016). Big data analytics for earth sciences: the EarthServer approach. International Journal of Digital Earth, 9(1), 3-29.

Baumann, P., Misev, D., Merticariu, V., Huu, B. P., Bell, B., Kuo, K. S., & Bayesics, L. L. C. (2018). Array Databases: Concepts, Standards, Implementations. Research Data Alliance (RDA) Working Group Report.

Bhangale, U. M., Kurte, K. R., Durbha, S. S., King, R. L., & Younan, N. H. (2016, July). Big data processing using hpc for remote sensing disaster data. In 2016 IEEE International Geoscience and Remote Sensing Symposium (IGARSS) (pp. 5894-5897). IEEE.

Blumenfeld J. (2019). Getting Petabytes to People: How EOSDIS Facilitates Earth Observing Data Discovery and Use. Available at: https://earthdata.nasa.gov/getting-petabytes-to-people-how-the-eosdis-facilitates-earth-observing-data-discovery-and-use , last accessed on May 1, 2019

Bouziane, H. L., Pérez, C., & Priol, T. (2008, August). A software component model with spatial and temporal compositions for grid infrastructures. In European Conference on Parallel Processing (pp. 698-708). Springer, Berlin, Heidelberg.

Callahan, S. P., Freire, J., Santos, E., Scheidegger, C. E., Silva, C. T., & Vo, H. T. (2006, June). VisTrails: visualization meets data management. In Proceedings of the 2006 ACM SIGMOD international conference on Management of data(pp. 745-747). ACM.

Camara, G., Assis, L. F., Ribeiro, G., Ferreira, K. R., Llapa, E., & Vinhas, L. (2016, October). Big earth observation data analytics: Matching requirements to system architectures. In Proceedings of the 5th ACM SIGSPATIAL international workshop on analytics for big geospatial data (pp. 1-6). ACM.

Chang, F., Dean, J., Ghemawat, S., Hsieh, W. C., Wallach, D. A., Burrows, M., ... & Gruber, R. E. (2008). Bigtable: A distributed storage system for structured data. *ACM Transactions on Computer Systems (TOCS)*, *26*(2), 4.

Chen, Q., Wang, L., & Shang, Z. (2008, December). MRGIS: A MapReduce-Enabled high performance workflow system for GIS. In 2008 IEEE Fourth International Conference on eScience (pp. 646-651). IEEE.

Chen, X. W., & Lin, X. (2014). Big data deep learning: challenges and perspectives. IEEE access, 2, 514-525.

Cheng, G., Zhou, P., & Han, J. (2016). Learning rotation-invariant convolutional neural networks for object detection in VHR optical remote sensing images. IEEE Transactions on Geoscience and Remote Sensing, 54(12), 7405-7415.

Clarke, K. C. (2003). Geocomputation’s future at the extremes: high performance computing and nanoclients. Parallel Computing, 29(10), 1281-1295.

Cudré-Mauroux, P., Kimura, H., Lim, K. T., Rogers, J., Simakov, R., Soroush, E., ... & DeWitt, D. (2009). A demonstration of SciDB: a science-oriented DBMS. Proceedings of the VLDB Endowment, 2(2), 1534-1537.

De Mauro, A., Greco, M., & Grimaldi, M. (2015, February). What is big data? A consensual definition and a review of key research topics. In AIP conference proceedings (Vol. 1644, No. 1, pp. 97-104). AIP.

Dean, J. (2016). Large-scale deep learning for building intelligent computer systems.

Dean, J., & Ghemawat, S. (2008). MapReduce: simplified data processing on large clusters. Communications of the ACM, 51(1), 107-113.

Desjardins, M. R., Hohl, A., Griffith, A., & Delmelle, E. (2018). A space–time parallel framework for fine-scale visualization of pollen levels across the Eastern United States. Cartography and Geographic Information Science, 1-13.

Ding, Y. and Densham, P., 1996. Spatial strategies for parallel spatial modelling. International Journal of Geographical Information Systems, 10 (6), 669–698. doi:10.1080/02693799608902104

Duffy, D., Spear, C., Bowen, M., Thompson, J., Hu, F., Yang, C., & Pierce, D. (2016, December). Emerging Cyber Infrastructure for NASA's Large-Scale Climate Data Analytics. In AGU Fall Meeting Abstracts.

Eldawy, A., & Mokbel, M. F. (2015, April). Spatialhadoop: A mapreduce framework for spatial data. In *2015 IEEE 31st international conference on Data Engineering* (pp. 1352-1363). IEEE.

Eldawy, A., & Mokbel, M. F. (2015, April). Spatialhadoop: A mapreduce framework for spatial data. In 2015 IEEE 31st international conference on Data Engineering (pp. 1352-1363). IEEE.

Eldawy, A., Mokbel, M. F., Alharthi, S., Alzaidy, A., Tarek, K., & Ghani, S. (2015, April). Shahed: A mapreduce-based system for querying and visualizing spatio-temporal satellite data. In 2015 IEEE 31st International Conference on Data Engineering(pp. 1585-1596). IEEE.

Engélinus, J., & Badard, T. (2018). Elcano: A Geospatial Big Data Processing System based on SparkSQL. In GISTAM (pp. 119-128).

Esri. (2013). GIS Tools for Hadoop. https://github.com/Esri/gis-tools-for-hadoop, last accessed on April 25, 2019

Fahmy, M. M., Elghandour, I., & Nagi, M. (2016, December). CoS-HDFS: co-locating geo-distributed spatial data in hadoop distributed file system. In Proceedings of the 3rd IEEE/ACM International Conference on Big Data Computing, Applications and Technologies (pp. 123-132). ACM.

Gabriel, E., Fagg, G. E., Bosilca, G., Angskun, T., Dongarra, J. J., Squyres, J. M., ... & Castain, R. H. (2004, September). Open MPI: Goals, concept, and design of a next generation MPI implementation. In European Parallel Virtual Machine/Message Passing Interface Users' Group Meeting(pp. 97-104). Springer, Berlin, Heidelberg.

Gong J., Wu H., Zhang T., Gui Z, Li Z., You L., Shen S., (2012). Geospatial Service Web: towards integrated cyberinfrastructure for GIScience. Geo-spatial Information Science, 15(2):73-84.

Goodchild, M. F. (2007). Citizens as sensors: the world of volunteered geography. GeoJournal, 69(4), 211-221

Google. (2019), Google BigQuery GIS, https://cloud.google.com/bigquery/docs/gis-intro, last accessed on April 25, 2019

Gorelick, N., Hancher, M., Dixon, M., Ilyushchenko, S., Thau, D., & Moore, R. (2017). Google Earth Engine: Planetary-scale geospatial analysis for everyone. *Remote Sensing of Environment*, *202*, 18-27.

Gropp, W., Lusk, E., Doss, N., & Skjellum, A. (1996). A high-performance, portable implementation of the MPI message passing interface standard. Parallel computing, 22(6), 789-828.

Guan, Q. (2009). pRPL: an open-source general-purpose parallel Raster Processing programming Library. Sigspatial Special, 1(1), 57-62.

Guan, Q., Zhang, T., & Clarke, K. C. (2006, December). GeoComputation in the grid computing age. In International Symposium on Web and Wireless Geographical Information Systems (pp. 237-246). Springer, Berlin, Heidelberg.

Gudivada, V. N., Baeza-Yates, R., & Raghavan, V. V. (2015). Big data: Promises and problems. Computer, (3), 20-23.

Gui, Z., Yu, M., Yang, C., Jiang, Y., Chen, S., Xia, J., Huang, Q., Liu, K., Li, Z., Hassan, M.A. and Jin, B., (2016). Developing Subdomain Allocation Algorithms Based on Spatial and Communicational Constraints to Accelerate Dust Storm Simulation. *PloS one*, 11(4), p.e0152250.

Guo, Z., Fox, G., & Zhou, M. (2012, May). Investigation of data locality in mapreduce. In Proceedings of the 2012 12th IEEE/ACM International Symposium on Cluster, Cloud and Grid Computing (ccgrid 2012) (pp. 419-426). IEEE Computer Society.

Guttman, A. (1984). R-trees: a dynamic index structure for spatial searching (Vol. 14, No. 2, pp. 47-57). ACM.

Hansen, M. C., Potapov, P. V., Moore, R., Hancher, M., Turubanova, S. A. A., Tyukavina, A., ... & Kommareddy, A. (2013). High-resolution global maps of 21st-century forest cover change. science, 342(6160), 850-853.

He, Z., Wu, C., Liu, G., Zheng, Z., & Tian, Y. (2015). Decomposition tree: a spatio-temporal indexing method for movement big data. Cluster Computing, 18(4), 1481-1492.

Hohl, A., D Griffith, A., Eppes, M. C., & Delmelle, E. (2018). Computationally Enabled 4D Visualizations Facilitate the Detection of Rock Fracture Patterns from Acoustic Emissions. Rock Mechanics and Rock Engineering, 51, 2733-2746.

Hohl, A., Delmelle, E. M., & Tang, W. (2015). SPATIOTEMPORAL DOMAIN DECOMPOSITION FOR MASSIVE PARALLEL COMPUTATION OF SPACE-TIME KERNEL DENSITY. ISPRS Annals of Photogrammetry, Remote Sensing & Spatial Information Sciences, 2(4).

Hu, F., Xia, G. S., Hu, J., & Zhang, L. (2015). Transferring deep convolutional neural networks for the scene classification of high-resolution remote sensing imagery. Remote Sensing, 7(11), 14680-14707.

Hu, F., Yang, C., Schnase, J. L., Duffy, D. Q., Xu, M., Bowen, M. K., ... & Song, W. (2018). ClimateSpark: An in-memory distributed computing framework for big climate data analytics. Computers & geosciences, 115, 154-166.

Huang, Z., Chen, Y., Wan, L., & Peng, X. (2017). GeoSpark SQL: An effective framework enabling spatial queries on spark. ISPRS International Journal of Geo-Information, 6(9), 285.

Hughes, J. N., Annex, A., Eichelberger, C. N., Fox, A., Hulbert, A., & Ronquest, M. (2015, May). Geomesa: a distributed architecture for spatio-temporal fusion. In Geospatial Informatics, Fusion, and Motion Video Analytics V (Vol. 9473, p. 94730F). International Society for Optics and Photonics.

Hull, D., Wolstencroft, K., Stevens, R., Goble, C., Pocock, M. R., Li, P., & Oinn, T. (2006). Taverna: a tool for building and running workflows of services. Nucleic acids research, 34(suppl_2), W729-W732.

Internet Live Stats, (2019), available at https://www.internetlivestats.com/twitter-statistics/, last accessed on May 3, 2019

Jaeger, E., Altintas, I., Zhang, J., Ludäscher, B., Pennington, D., & Michener, W. (2005, June). A Scientific Workflow Approach to Distributed Geospatial Data Processing using Web Services. In SSDBM (Vol. 3, No. 42, pp. 87-90).

Jia, Y., Shelhamer, E., Donahue, J., Karayev, S., Long, J., Girshick, R., ... & Darrell, T. (2014, November). Caffe: Convolutional architecture for fast feature embedding. In Proceedings of the 22nd ACM international conference on Multimedia (pp. 675-678). ACM.

Katal, A., Wazid, M., & Goudar, R. H. (2013, August). Big data: issues, challenges, tools and good practices. In *2013 Sixth international conference on contemporary computing (IC3)* (pp. 404-409). IEEE.

Kini, A., & Emanuele, R. (2014). Geotrellis: Adding geospatial capabilities to spark. Spark Summit.

Kussul, N., Lavreniuk, M., Skakun, S., & Shelestov, A. (2017). Deep learning classification of land cover and crop types using remote sensing data. IEEE Geoscience and Remote Sensing Letters, 14(5), 778-782.

LeCun, Y., Bengio, Y., & Hinton, G. (2015). Deep learning. nature, 521(7553), 436.

Lee, K., & Kim, K. (2018, July). Geo-Based Image Analysis System Supporting OGC-WPS Standard on Open PaaS Cloud Platform. In IGARSS 2018-2018 IEEE International Geoscience and Remote Sensing Symposium (pp. 5262-5265). IEEE.

Li Z., Hodgson M., Li W.  (2018) A general-purpose framework for large-scale Lidar data processing, International Journal of Digital Earth, 11(1), 26-47

Li Z., Huang Q., Jiang Y., Hu F. (2019) , SOVAS: A Scalable Online Visual Analytic System for Big Climate Data Analysis, International Journal of Geographic Information Science, DOI: 10.1080/13658816.2019.1605073

Li Z., Yang C., Yu M., Liu K., Sun M.(2015) Enabling Big Geoscience Data Analytics with a Cloud-based, MapReduce-enabled and Service-oriented Workflow Framework, *PloS one*, 10(3), e0116781.

Li Z., Yang, C., Huang, Q., Liu K., Sun, M., Xia, J., (2017a). Building Model as a Service for Supporting Geosciences,*Computers, Environment and Urban Systems*. 61(B), 141-152

Li Z., Huang Q., Carbone G., Hu F. (2017b) A High Performance Query Analytical Framework for Supporting Data-intensive Climate Studies, Computers, Environment and Urban Systems, 62(3), 210-221

Li, J., Jiang, Y., Yang, C., Huang, Q., & Rice, M. (2013). Visualizing 3D/4D environmental data using many-core graphics processing units (GPUs) and multi-core central processing units (CPUs). Computers & Geosciences, 59, 78-89.

Li, Z. (2015). Optimizing geospatial cyberinfrastructure to improve the computing capability for climate studies (Doctoral dissertation), available at: http://ebot.gmu.edu/bitstream/handle/1920/9630/Li_gmu_0883E_10873.pdf?sequence=1&isAllowed=y

Li, Z., Yang, C. P., Wu, H., Li, W., & Miao, L. (2011). An optimized framework for seamlessly integrating OGC Web Services to support geospatial sciences. International Journal of Geographical Information Science, 25(4), 595-613.

Li, Z., Hu, F., Schnase, J. L., Duffy, D. Q., Lee, T., Bowen, M. K., & Yang, C. (2017c). A spatiotemporal indexing approach for efficient processing of big array-based climate data with MapReduce. International Journal of Geographical Information Science, 31(1), 17-35.

Li, Z., Yang, C., Liu, K., Hu, F., & Jin, B. (2016). Automatic Scaling Hadoop in the Cloud for Efficient Process of Big Geospatial Data. *ISPRS International Journal of Geo-Information*, 5(10), 173.

Ling, F., & Foody, G. M. (2019). Super-resolution land cover mapping by deep learning. Remote Sensing Letters, 10(6), 598-606.

Ma, L.; Li, M.; Ma, X.; Cheng, L.; Du, P.; Liu, Y. A review of supervised object-based land-cover image classification. ISPRS J. Photogramm. Remote Sens. 2017, 130, 277–293.

Manyika, J., Chui, M., Brown, B., Bughin, J., Dobbs, R., Roxburgh, C. and Byers, A. H. (2011). Big data: The next frontier for innovation, competition, and productivity, Big Data: The Next Frontier for Innovation, Competition & Productivity, pp. 1-143, Available online at https://bigdatawg.nist.gov/pdf/MGI_big_data_full_report.pdf

Marciniec, M. (2017), Observing World Tweeting Tendencies in Real-time, available at https://codete.com/blog/observing-world-tweeting-tendencies-in-real-time-part-2, last accessed on May 3, 2019

Minsky, M. (1961). Steps toward artificial intelligence. Proceedings of the IRE, 49(1), 8-30.

Murphy JM, Sexton DM, Barnett DN, Jones GS, Webb MJ, et al. (2004) Quantification of modelling uncertainties in a large ensemble of climate change simulations. Nature 430: 768-772

Nvidia, C. U. D. A. (2011). Nvidia cuda c programming guide. *Nvidia Corporation*, *120*(18), 8.

OGC (2017), OGC announces a new standard that improves the way information is referenced to the earth

O'Leary, D. E. (2013). BIG DATA', THE 'INTERNET OF THINGS'AND THE 'INTERNET OF SIGNS. Intelligent Systems in Accounting, Finance and Management, 20(1), 53-65.

Ooi, B. C. (1987). Spatial kd-tree: A data structure for geographic database. In Datenbanksysteme in Büro, Technik und Wissenschaft (pp. 247-258). Springer, Berlin, Heidelberg.

Ooi, B. C., Tan, K. L., Wang, S., Wang, W., Cai, Q., Chen, G., ... & Xie, Z. (2015, October). SINGA: A distributed deep learning platform.

In *Proceedings of the 23rd ACM international conference on Multimedia* (pp. 685-688). ACM.

Planthaber, G., Stonebraker, M., & Frew, J. (2012, November). EarthDB: scalable analysis of MODIS data using SciDB. In Proceedings of the 1st ACM SIGSPATIAL International Workshop on Analytics for Big Geospatial Data (pp. 11-19). ACM.

Ramsey, P. (2005). Postgis manual. Refractions Research Inc, 17.

Rienecker, M. M., Suarez, M. J., Gelaro, R., Todling, R., Bacmeister, J., Liu, E., ... & Bloom, S. (2011). MERRA: NASA's modern-era retrospective analysis for research and applications. Journal of climate, 24(14), 3624-3648.

Robinson , 2012, The Storage and Transfer Challenges of Big Data, http://sloanreview.mit.edu/article/the-storage-and-transfer-challenges-of-big-data/ (Last accessed Nov. 25, 2015)

Sabeur, Z, Gibb, R., Purss, M., 2019.  Discrete Global Grid Systems SWG, Available at http://www.opengeospatial.org/projects/groups/dggsswg. Last accessed on March 13, 2019

Samet, H. (1984). The quadtree and related hierarchical data structures. ACM Computing Surveys (CSUR), 16(2), 187-260.

Shook, E., Hodgson, M. E., Wang, S., Behzad, B., Soltani, K., Hiscox, A., & Ajayakumar, J. (2016). Parallel cartographic modeling: a methodology for parallelizing spatial data processing. International Journal of Geographical Information Science, 30(12), 2355-2376.

Shvachko, K., Kuang, H., Radia, S., & Chansler, R. (2010, May). The hadoop distributed file system. In *MSST* (Vol. 10, pp. 1-10).

Stonebraker, M. (1986). The case for shared nothing. *IEEE Database Eng. Bull.*, *9*(1), 4-9.

Tan, X., Di, L., Deng, M., Fu, J., Shao, G., Gao, M., ... & Jin, B. (2015). Building an elastic parallel OGC web processing service on a cloud-based cluster: A case study of remote sensing data processing service. Sustainability, 7(10), 14245-14258.

Tang, W., Feng, W., & Jia, M. (2015). Massively parallel spatial point pattern analysis: Ripley’s K function accelerated using graphics processing units. International Journal of Geographical Information Science, 29(3), 412-439.

Taylor, Ian, Ian Wang, Matthew Shields, and Shalil Majithia. "Distributed computing with Triana on the Grid." Concurrency and Computation: Practice and Experience 17, no. 9 (2005): 1197-1214.

Taylor, R. C. (2010, December). An overview of the Hadoop/MapReduce/HBase framework and its current applications in bioinformatics. In BMC bioinformatics (Vol. 11, No. 12, p. S1). BioMed Central.

Thain, D., Tannenbaum, T., & Livny, M. (2005). Distributed computing in practice: the Condor experience. Concurrency and computation: practice and experience, 17(2 - 4), 323-356.

Tschauner, H., & Salinas, V. S. (2006, April). Stratigraphic modeling and 3D spatial analysis using photogrammetry and octree spatial decomposition. In Digital Discovery. Exploring New Frontiers in Human Heritage. CAA2006. Computer Applications and Quantitative Methods in Archaeology. Proceedings of the 34th Conference, Fargo, United States(pp. 257-270).

Unat, D., Dubey, A., Hoefler, T., Shalf, J., Abraham, M., Bianco, M., ... & Fuerlinger, K. (2017). Trends in data locality abstractions for HPC systems. IEEE Transactions on Parallel and Distributed Systems, 28(10), 3007-3020.

VoPham, T., Hart, J. E., Laden, F., & Chiang, Y. Y. (2018). Emerging trends in geospatial artificial intelligence (geoAI): potential applications for environmental epidemiology. Environmental Health, 17(1), 40.

Vora, M. N. (2011, December). Hadoop-HBase for large-scale data. In *Proceedings of 2011 International Conference on Computer Science and Network Technology* (Vol. 1, pp. 601-605). IEEE.

Wang, F., Aji, A., Liu, Q., & Saltz, J. (2011). Hadoop-GIS: A high performance spatial query system for analytical medical imaging with MapReduce. Center for Comprehensive Informatics, Technical Report. Available at: http://www3. cs. stonybrook. edu/~ fuswang/papers/CCI-TR-2011–3. pdf (access 21 September 2015)(online).

Wang, L., Chen, B., & Liu, Y. (2013, June). Distributed storage and index of vector spatial data based on HBase. In *2013 21st international conference on geoinformatics* (pp. 1-5). IEEE.

Wang, L., Chen, B., & Liu, Y. (2013, June). Distributed storage and index of vector spatial data based on HBase. In 2013 21st international conference on geoinformatics (pp. 1-5). IEEE.

Wang, S. (2008, November). GISolve toolkit: advancing GIS through cyberinfrastructure. In Proceedings of the 16th ACM SIGSPATIAL international conference on Advances in geographic information systems (p. 83). ACM.

Wang, S. (2010). A CyberGIS framework for the synthesis of cyberinfrastructure, GIS, and spatial analysis. Annals of the Association of American Geographers, 100(3), 535-557.

Wang, S., & Armstrong, M. P. (2003). A quadtree approach to domain decomposition for spatial interpolation in grid computing environments. Parallel Computing, 29(10), 1481-1504.

Wang, S., Anselin, L., Bhaduri, B., Crosby, C., Goodchild, M. F., Liu, Y., & Nyerges, T. L. (2013). CyberGIS software: a synthetic review and integration roadmap. International Journal of Geographical Information Science, 27(11), 2122-2145.

Ward and Barker (2013). School of Computer Science, University of St Andrews, Undefined By Data: A Survey of Big Data Definitions UK

Whitby, M. A., Fecher, R., & Bennight, C. (2017, August). Geowave: Utilizing distributed key-value stores for multidimensional data. In *International Symposium on Spatial and Temporal Databases* (pp. 105-122). Springer, Cham.

Widlund, O. B. (2009). Accomodating irregular subdomains in domain decomposition theory. In Domain decomposition methods in science and engineering XVIII (pp. 87-98). Springer, Berlin, Heidelberg.

Wu, D., Liu, S., Zhang, L., Terpenny, J., Gao, R. X., Kurfess, T., & Guzzo, J. A. (2017). A fog computing-based framework for process monitoring and prognosis in cyber-manufacturing. Journal of Manufacturing Systems, 43, 25-34.

Wulder, M. A., White, J. C., Loveland, T. R., Woodcock, C. E., Belward, A. S., Cohen, W. B., ... & Roy, D. P. (2016). The global Landsat archive: Status, consolidation, and direction. Remote Sensing of Environment, 185, 271-283.

Xia, J., Yang, C., Gui, Z., Liu, K., & Li, Z. (2014). Optimizing an index with spatiotemporal patterns to support GEOSS Clearinghouse. International Journal of Geographical Information Science, 28(7), 1459-1481.

Xie, D., Li, F., Yao, B., Li, G., Zhou, L., & Guo, M. (2016, June). Simba: Efficient in-memory spatial analytics. In Proceedings of the 2016 International Conference on Management of Data (pp. 1071-1085). ACM.

Yang, C., Huang, Q., Li, Z., Liu K., Hu, F. (2017). Big Data and cloud computing: innovation opportunities and challenges, *International Journal of Digital Earth*, 10(1),1-41.

Yin, D., Liu, Y., Padmanabhan, A., Terstriep, J., Rush, J., & Wang, S. (2017, July). A CyberGIS-Jupyter framework for geospatial analytics at scale. In Proceedings of the Practice and Experience in Advanced Research Computing 2017 on Sustainability, Success and Impact (p. 18). ACM.

Yin, J., Foran, A., & Wang, J. (2013, October). Dl-mpi: Enabling data locality computation for mpi-based data-intensive applications. In *2013 IEEE International Conference on Big Data* (pp. 506-511). IEEE.

Yoon, G., & Lee, K. (2015). WPS-based Satellite Image Processing onWeb Framework and Cloud Computing Environment. Korean Journal of Remote Sensing, 31(6), 561-570.

Yu, J., Wu, J., & Sarwat, M. (2015, November). Geospark: A cluster computing framework for processing large-scale spatial data. In Proceedings of the 23rd SIGSPATIAL International Conference on Advances in Geographic Information Systems(p. 70). ACM.

Yue, P., Gong, J., & Di, L. (2010). Augmenting geospatial data provenance through metadata tracking in geospatial service chaining. Computers & Geosciences, 36(3), 270-281.

Zaharia, M., Xin, R. S., Wendell, P., Das, T., Armbrust, M., Dave, A., ... & Ghodsi, A. (2016). Apache spark: a unified engine for big data processing. *Communications of the ACM*, *59*(11), 56-65.

Zhang, C., Di, L., Sun, Z., Eugene, G. Y., Hu, L., Lin, L., ... & Rahman, M. S. (2017, August). Integrating OGC Web Processing Service with cloud computing environment for Earth Observation data. In 2017 6th International Conference on Agro-Geoinformatics (pp. 1-4). IEEE.

Zhang, C.; Sargent, I.; Pan, X.; Li, H.; Gardiner, A.; Hare, J.; Atkinson, P.M. An object-based convolutional neural network (OCNN) for urban land use classification. Remote Sens. Environ. 2018, 216, 57–70.

Zhang, J., Pennington, D. D., & Michener, W. K. (2006, May). Automatic transformation from geospatial conceptual workflow to executable workflow using GRASS GIS command line modules in Kepler.

In International Conference on Computational Science (pp. 912-919). Springer, Berlin, Heidelberg.

Zhang, X., Song, W., & Liu, L. (2014, June). An implementation approach to store GIS spatial data on NoSQL database. In *2014 22nd international conference on geoinformatics* (pp. 1-5). IEEE.

Zhao, L., Chen, L., Ranjan, R., Choo, K. K. R., & He, J. (2016). Geographical information system parallelization for spatial big data processing: a review. Cluster Computing, 19(1), 139-152.

Zheng, M., Tang, W., Lan, Y., Zhao, X., Jia, M., Allan, C., & Trettin, C. (2018). Parallel Generation of Very High Resolution Digital Elevation Models: High-Performance Computing for Big Spatial Data Analysis. In Big Data in Engineering Applications (pp. 21-39). Springer, Singapore.

Zikopoulos, P., & Eaton, C. (2011). Understanding big data: Analytics for enterprise class hadoop and streaming data. McGraw-Hill Osborne Media.

Zikopoulos, P., Parasuraman, K., Deutsch, T., Giles, J., & Corrigan, D. (2012). Harness the power of big data The IBM big data platform. McGraw Hill Professional.